\documentstyle[epsfig,twocolumn,aps]{revtex}
\begin{document}
\draft
\title{Scattering of short laser pulses from trapped fermions}
\author{T. Wong, \"{O}zg\"{u}r M\"{u}stecapl{\i}o\={g}lu, and L. You}
\address{School of Physics, Georgia Institute of Technology,
Atlanta, Georgia 30332-430, USA}
\author{Maciej Lewenstein}
\address{Institut f\"ur Theoretische Physik, Universit\"at
Hannover, D-30163 Hannover, Germany}
\date{\today}
\maketitle

\begin{abstract}
We investigate the scattering of intense short laser pulses off
trapped cold fermionic atoms. We discuss the sensitivity of the
scattered light to the quantum statistics of the atoms. The
temperature dependence of the scattered light spectrum is
calculated. Comparisons are made with a system of classical atoms
who obey Maxwell-Boltzmann statistics. We find the total
scattering increases as the fermions become cooler but eventually
tails off at very low temperatures (far below the Fermi
temperature). At these low temperatures the fermionic degeneracy
plays an important role in the scattering as it inhibits
spontaneous emission into occupied energy levels below the Fermi
surface. We demonstrate temperature dependent qualitative changes
in the differential and total spectrum can be utilized to probe
quantum degeneracy of trapped Fermi gas when the total number of
atoms are sufficiently large $(\geq 10^6)$. At smaller number of
atoms, incoherent scattering dominates and it displays weak
temperature dependence.
\end{abstract}


%\pacs{42.50.Fx,42.50.-p,32.80.-t}

\narrowtext

\section{Introduction}

Rapid advances of trapping and cooling of alkali atoms \cite{Chu} have led
to the recent dramatic achievement of Bose-Einstein condensation (BEC) \cite
{jila,mit,rice,edwards}. Focusing these highly successful trapping and
cooling methods onto fermions instead of bosons offers additional rich
opportunities for studying degenerate fermionic atomic gases. The
description of these magnetically trapped Fermi gases presents additional
simplification as the Pauli exclusion principle forbids the unsuppressed low
energy s-wave collisions among atoms in the same hyperfine state. Thus these
dilute trapped atoms behave very close to the ideal Fermi gas so ubiquitous
in physics textbooks for decades. Recent highlights from several leading
experimental groups have indicated that we are at the edge of being able to
explore quantum degenerate Fermi gas \cite{Jinn,ens} inside laboratories.

Once such a degenerate gas is achieved, how do we perform
diagnostic measurements on its properties ? A standard method in
atomic physics is to probe the gas with light scattering. This is
the problem to be addressed in our paper. These spectroscopic
light scattering methods have already been suggested for BEC in
both the weak and strong field scattering regimes \cite {ly,lbec}.
These early theoretical investigations had limited impact on BEC
experimental observations so far, partly because the resonant
photon-atom interaction becomes a complicated many body problem
when a condensate is involved. The recent dramatic demonstration
of low group velocity of light propagation inside a condensate
\cite{hau}, the Bragg scattering experiment \cite{bragg}, and the
surprising observation of super-radiant collective spontaneous
emission from MIT \cite{ship} all call for more detailed
applications of quantum field theory of photons interacting with
atoms. In the last several years, light scattering off Fermi
degenerate atoms have already been discussed by several groups;
some of these investigations focused on the case of a distribution
of ideal atoms obeying Fermi-Dirac statistics \cite{fdd}; while
others considered more exotic state for a Cooper paired fermionic
condensate \cite{BCS}. One notable feature for a Fermi degenerate
gas is that the Pauli exclusion principle blocks (inhibits)
scattering events for atoms into already occupied states
\cite{zoller}. This paper aims at detailed quantitative
information of light scattering from trapped Fermi gases within
parameter regimes of current/future experiments, specifically
results explicitly demonstrating temperature and quantum
statistical effects are sought.

In a previous paper involving two of us (LY and ML) an optical method for
detection of the properties of BEC was proposed \cite{short}. This detection
involves the limiting case of scattering short but intense laser pulses from
a system of cooled bosonic atoms in a trap. In particular, the case of laser
pulses with areas of 2$\pi K$ was investigated. (The pulse area to be
defined later is proportional to the integral of the slowly varying envelope
of the electric field multiplied by the atomic dipole moment). It was shown
that such pulses mainly cause cyclic Rabi oscillations for atoms in their
excited and ground states. Thus to zero-th order involves no photon
scattering (except the stimulated interaction with the driving field).
However, atoms can spontaneously emit photons while in their excited states,
therefore, the previous zero-th order picture involving coherent evolution
of all atoms can be interrupted by spontaneous emissions of individual
atoms. Thus scattered light from these emissions can be collected and their
properties reflect to a certain degree the properties of the trapped atoms.
It was shown that for the case of BEC, above the critical temperature, $T_c$%
, the coherent scattering is very weak and is predominately in the forward
direction due to phase matching effects. However, below the critical
temperature the number of scattered photons increases dramatically and the
coherent scattering occurs within a solid angle determined by the size of
the condensate. For sufficiently short $2\pi K$-pulses the system is
preserved even below $T_c$ so that this method can be potentially used as a
non-destructive probe of BEC.

This paper presents a continuation of our investigations to the
case of a system of trapped {\it fermionic} atoms. The ideal gas
of a Fermi-Dirac distribution is considered in this paper. The
static properties of harmonic trapped ideal fermionic atoms as
considered by several groups previously will be used as inputs
\cite{fddoo}. The more subtle case involving a weak attractive
interactions between atoms could potentially develop into a BCS
type Cooper paired condensate \cite{bcsp}, and whose pulsed light
scattering properties \cite{BCS} will be explored in a future
publication. In the present study involving ideal non-interacting
fermionic atoms, we do not expect to see a dramatic change in the
spectrum as one cools the gas since no phase transition occurs in
the trapped gases even at zero temperature. However, with a finite
number of atoms (say $1$ million) and cooled far below the Fermi
temperature \cite{Jinn} one does expect that the quantum
statistics should play a role in the spectrum. Indeed, we
demonstrate that when number of atoms are sufficiently large such
that coherent scattering dominates, spectrum of scattered light
exhibits qualitative changes in its temperature dependence, a
signature potentially useful to probe quantum degeneracy of the
system.

The organization of the paper is as follows; first we review the
formulation as was presented previously \cite{ly} for trapped
bosons. In the second quantized form, the only difference lies at
the commutation relations between atomic operators. We then
calculate the spectra of scattered light taking into account the
fermionic nature of the atomic operators. In Sect. II, we present
the formulation of our model system. The near resonant laser pulse
is assumed to be intense so that its interaction with trapped
atoms is taken to be the zeroth order coherent process for which
exact analytical solutions for the dynamics of atomic operators
are given. In Sect. III, the interaction of weak scattered light
with trapped atoms is treated quantum mechanically with
perturbation method. The zeroth order solution of atomic operators
from Sect. II is used to evaluate the dynamics of scattered
photons. In Section IV, our new results for the total and
differential spectrums of scattered photons are then displayed
numerically. A discussion section follows where both the angular
and spectral distributions are presented. We also find the total
number of atoms scattered for typical experimental parameter sets.
Finally we conclude with a summary of the essential physics
learned from this investigation. In the appendix we show how we
can re-write the expression for the spectrum of the coherently
scattered light in a form that is suitable for numerical
calculation.

As will become obvious by the introduction of many newly derived analytic
formulae, the trapped fermion case is numerically far more difficult to
calculate than the previous bosonic one. At low temperatures the fermionic
system consists of many stacked energy levels \cite{fddoo} whereas the
bosonic system nicely condenses to a few of the lowest energy levels. Also
the numerical methods for infinite series summations used previously for
bosons are only applicable for the high temperature regime of a Fermi gas.
New methods are therefore developed to cope with the additional difficulties
of the fermionic system.

The first few sections of this paper follows closely to the BEC situation.
Initially the treatment is formally the same for both bosons and fermions,
the only difference being the commutation relations. However, these
differences become significant later especially with the calculation of the
spectrum.

\section{The model}

We consider a system consisting of $N$ fermionic atoms confined in a trap
interacting with light. Let us write down its Hamiltonian in the Fock
representation and in the second quantized form \cite{ly} :
\begin{eqnarray}
{\cal H}&=&\sum_{\vec n} \omega_{\vec n}^{g} g_{\vec n}^{\dag} g_{\vec n}+
\sum_{\vec m} (\omega_{\vec m}^{e}+\omega_0) {\vec e}_{\vec m}^{\;\dag}{\vec %
e}_{\vec m}  \nonumber \\
&+&\sum_{\vec n,\vec m}\sum_{\mu}\int d{\vec k}\varrho(k) \big[\eta_{\vec n%
\vec m}({\vec k})g_{\vec n}^{\dag}a^{\dag}_{{\vec k}\mu} {\vec e}_{\vec m%
}\!\cdot{\vec \epsilon}_{{\vec k}\mu}+{\rm h.c.}\big]  \nonumber \\
&+&\sum_{\mu}\int d{\vec k}\ ck\; a^{\dag}_{{\vec k}\mu}a_{{\vec k}\mu} .
\label{H}
\end{eqnarray}
where the Franck-Condon factors are
\begin{equation}
\eta_{\vec n\vec m}({\vec k})= \langle{g},{\vec n}|e^{-i{\vec k}\cdot{\vec R}%
} |{e},{\vec m}\rangle  \label{dnm}
\end{equation}
with $\vec R$ denoting the position operator of the atom (its nucleus). We
have used atomic units (unless otherwise stated), the rotating wave and the
dipole approximation. The atomic annihilation and creation operators for the
$\vec n$-th state of the center of mass motion of the atoms in the trap are
denoted by $g_{\vec n}$ and $g_{\vec n}^{\dag}$ respectively. Since these
operators are associated with the atoms in the ground electronic state, for
the case of a spherically symmetric harmonic trap potential, $\vec n$ has
three components $(n_x,n_y,n_z)$ and energy $\omega_{\vec n}^{g}=\omega_{%
{\rm t}}(n_x+n_y+n_z)$ where $\omega_{{\rm t}}$ is the trap frequency. The
size of the trap is related to the size of the ground state of the trap
potential, $a=\sqrt{1/2M\omega_{{\rm t}}}$ ($\hbar=1$). The atomic
annihilation and creation operators in the excited state ${\vec e}_{\vec m}$
and ${\vec e}_{\vec m}^{\;\dag}$ may experience different trap potential
from that of the ground state. However, from our analysis of short pulse
scattering, we find that the particular shape of the excited state potential
is unimportant since atoms only spend a very short period of time in the
excited state \cite{ly}. The electronic transition occurs at the frequency $%
\omega_0$. Since we are treating an $s$-state to a $p$-state transition the
excited state operators are vectors ${\vec e}_{\vec m}$ and ${\vec e}_{\vec m%
}^{\;\dag}$. Annihilation and creation operators for photons of momentum $%
\vec k$ and linear polarization ${\vec \epsilon}_{{\vec k}\mu}$ ($\mu=1,2$)
are denoted by $a_{{\vec k}\mu}$ and $a^{\dag}_{{\vec k}\mu}$. All atomic
operators obey standard fermionic anti-commutation relations. $\varrho(k)$
is a slowly varying coupling which is dependent on $k$. Its relation to the
natural linewidth is $\gamma=({8\pi^2k_0^2}/{3c})\big|\varrho(k_0)\big|^2$,
with $k_0=\omega_0/c$. For notational convenience, we will suppress the
indices ${g,e}$ for the internal states. The convention being that the
indices $\vec n$, $\vec n^{\prime}$ represents the center of mass states in
the electronic ground potential whereas $\vec m$, $\vec m^{\prime}$ denotes
the center of mass states in the excited state potentials.

Note that the strong resonant atomic dipole-dipole interaction resulting
from the exchange of transverse photons is included in Eq. (\ref{H}).

If the system is driven by a coherent laser pulse then we may
neglect spontaneous emission effects during the pulse if the pulse
is sufficiently short and intense. We can then safely substitute
the electric field operator entering the interaction Hamiltonian
in Eq.~(\ref{H}) by a $c$-number. The pulses we intend to use
should have duration $\tau_L\leq 300$ ps or shorter, i.e. width
$\gamma_L=1/\tau_L\simeq3\times 10^9-10^{11}$ Hz. A first estimate
shows that $\gamma_L\gg\gamma\simeq 2.5$ MHz, so that spontaneous
emission may be neglected during the interaction time of the pulse
with the atoms. The current estimate for single excitation
spontaneous emission is a far better one than in the bosonic case
as there is no Bose enhancement. However, the fermionic nature may
come into play at very low temperatures where the effective
spontaneous emission rate will be greatly reduced due to
suppression by the Fermi sea of ground levels. Therefore it is a
good approximation to assume that the effects of dissipative
spontaneous emission and dispersive dipole-dipole interactions are
small compared to the coherent driving laser during the
interaction between the atoms and the laser pulse. We can then
replace the product of the electric field operator ($\vec {{\cal
E}}^{\,(+)}$) and the absolute value of the electronic transition
dipole moment ($d$) by
\begin{equation}
d{\vec {{\cal E}}}^{\,(+)}\to\frac{\Omega}{2} \sum_{\mu}\int d{\vec k}\;{%
\vec \varrho}({\vec k},\mu) e^{i{\vec k}\cdot{\vec R}-ickt}.  \label{sub}
\end{equation}
The envelope of the laser pulse is defined by ${\vec \varrho}({\vec k},\mu)$%
. $\Omega$ is the peak Rabi frequency of the laser pulse. Using the
assumption that the pulse is a plane wave packet moving in the ${\vec k}_L$
direction with a central frequency $\omega_L$ and a linear polarization ${%
\vec \epsilon}_L$, we obtain
\begin{equation}
d{\vec {{\cal E}}}^{(+)}\to\frac{\Omega}{2}{\vec \epsilon}_L {\cal T}\big[%
\gamma_L(t-{\vec k}_L\cdot{\vec R}/\omega_L)\big] e^{i{\vec k}_L\cdot{\vec R}%
-i\omega_Lt}.  \label{pulse}
\end{equation}
The time dependent profile of the pulse, ${\cal T}(\gamma_Lt)$ is chosen to
be real and we assume a gaussian shape with a peak equal to one at $t=0$.

In order for Eq.~(\ref{pulse}) to be valid, we need ${\vec\varrho}({\vec k}%
,\mu)$ to vary in momentum of the order of $\gamma_L/c\simeq 10-300$ m$^{-1}$%
. However, the Franck-Condon factors $\eta_{\vec n\vec m}({\vec
k})$ vary on the scale of $\delta k$, on the order of $1/a\simeq
10^5\ {\rm m}^{-1}$ for low $n$. For higher $n$'s, $\delta k$
scales as $1/\sqrt{n}$, so it becomes $\sim 10^3 $ m$^{-1}$ for
the highest energy levels that are still available in the trap.
Thus, we have $\delta k\gg \gamma_L/c$ for all $\gamma_L$, so we
may validly replace ${\vec k}$ by ${\vec k}_L$ inside $\eta_{\vec n\vec m}({%
\vec k}_L)$. Inserting this substitution into Eq. (\ref{sub}) and Eq. (\ref
{pulse}), Eq.~(\ref{H}) can be written as
\begin{eqnarray}
{\cal H}&=&\sum_{\vec n} \omega_{\vec n}^{g} g_{\vec n}^{\dag} g_{\vec n}
+\sum_{\vec m} (\omega_{\vec m}^{e}+\omega_0) {\vec e}_{\vec m}^{\;\dag} {%
\vec e}_{\vec m}  \nonumber \\
&+&\frac{\Omega}{2}{\cal T}(\gamma_Lt)\Big[\exp(i\omega_Lt) \sum_{n}g_{\vec n%
}^{\dag}{\vec\epsilon}_L\cdot{\vec f}_{n}+{\rm h.c.}\Big],  \label{H2}
\end{eqnarray}
where we have re-written the operators in terms of annihilation and creation
operators of wave packets of excited states which originate from the $\vec n$%
-th state of the ground state potential,
\begin{equation}
{\vec f}_{\vec n} =\sum_{\vec m} \eta_{\vec n\vec m}({\vec k}_L) {\vec e}_{%
\vec m}.  \label{f}
\end{equation}
These annihilation operators and their conjugate creation ones also obey the
standard fermionic anti-commutation relations, i.e. $\{f_{\vec n}^{q},f_{%
\vec n^{\prime}}^{q^{\prime}\dag}\} =\delta_{\vec n\vec n^{\prime}}%
\delta_{qq^{\prime}}$, with $q,q^{\prime}=x,y,z$ enumerating the components
of the vectors $\vec f_{\vec n}$ and $\vec f_{\vec n^{\prime}}^{\dag}$.
Their energies $\omega_{\vec m}$ also vary very slowly for their
corresponding states and therefore for each of the wave packets ${\vec f}_{%
\vec n}$, ${\vec f}_{\vec n}^{\;\dag}$, their energy can be approximated by $%
\omega_{\vec n}^{g}+\omega_0+k_L^2/(2M)$. This assumes that the atomic
wavepackets in the excited state potential will not experience much coherent
oscillation or diffusion (i.e. are in a sense frozen in shape) within the
duration of the laser pulse ($\tau_L\ll 1/\omega_{{\rm t}}$).

The Heisenberg equations that follow from the Hamiltonian (\ref{H2}) now
becomes linear. Thus at resonance, $\omega_{L}\approx \omega_0+k^2_L/(2M)$,
and in the rotating frame in which $g_{\vec n}\to e^{-i\omega_{\vec n%
}^{g}t}g_{\vec n}$, ${\vec f}_{\vec n}\to e^{-i(\omega^g_{\vec n%
}+\omega_L)t}f_{\vec n}$, they become
\begin{eqnarray}
{\dot g_{\vec n}}(t)&=&-i\frac{\Omega}{2} {\cal T}(\gamma_Lt){\vec\epsilon}%
_L\cdot{\vec f}_{\vec n}(t), \\
{\vec\epsilon}_L\cdot{\dot{{\vec f}_{\vec n}\,}}(t)&=&-i\frac{\Omega}{2}
{\cal T}(\gamma_Lt)g_{\vec n}(t).
\end{eqnarray}
They can be easily solved analytically for any pulse envelope
\begin{eqnarray}
g_{\vec n}(t)&=& g_{\vec n}(-\infty)\cos\big[A(t)\big]  \nonumber \\
&&-i{\vec\epsilon}_L \cdot{\vec f}_{\vec n}(-\infty)\sin\big[A(t)\big],
\label{eqs8i} \\
{\vec\epsilon}_L\cdot{\vec f}_{\vec n}(t)&=& -ig_{\vec n}(-\infty)\sin\big[%
A(t)\big]  \nonumber \\
&&+{\vec\epsilon}_L\cdot{\vec f}_{\vec n}(-\infty)\cos\big[A(t)\big],
\label{eqs8}
\end{eqnarray}
with the pulse area
\begin{equation}
A(t)=\frac{\Omega}{2}\int_{-\infty}^t{\cal T}(\gamma_Lt^{\prime})dt^{\prime}.
\end{equation}
Thus we see that each of the $\vec n$ levels of the ground state oscillator
(when populated) creates an independent wave packet ${\vec f}_{\vec n}$
which is a superposition of the excited state wavefunctions. The population
oscillates coherently between the $\vec n$-th ground state and the
corresponding excited state wave packet. We can view the behavior of the
system as a set of independent two-level atoms coherently driven by the
laser pulse. By using a pulse whose area is a multiple of $2\pi$ the system
will be left in the same state after the duration of the pulse. Clearly, as $%
|\vec n|$ increases, the approximations become less valid, but they should
hold very well for the lowest $10^6$ available states of the ground state
potential.

The linear relations ${\vec f}_{\vec n}$ and ${\vec e}_{\vec m}$ remain the
same as before for bosonic atoms \cite{ly} and their inverse can be easily
derived using the sum rules
\begin{eqnarray}
&&\sum_{\vec n}\big[\eta_{\vec n\vec m}({\vec k}_L)\big]^{*} \eta_{\vec n%
\vec m^{\prime}}({\vec k}_L)=\delta_{\vec m\vec m^{\prime}}, \\
&&\sum_{\vec m}\big[\eta_{\vec n\vec m}({\vec k}_L)\big]^{*} \eta_{\vec n%
^{\prime}\vec m}({\vec k}_L)=\delta_{\vec n\vec n^{\prime}}.  \label{sumr1}
\end{eqnarray}
They will allow us to express solutions for $g_{\vec n}(t)$ and $\vec e_{%
\vec m}(t)$ as given above in Eqs. (\ref{eqs8i}) and (\ref{eqs8}) (and their
conjugates) uniquely in terms of $g_{\vec n}(-\infty)$, $\vec e_{\vec m%
}(-\infty)$, etc.

Since spontaneous emission rate $\gamma$ from a dipole allowed excite states
is non-zero, we do in practice have photons scattered from the atoms. The
detailed discussions were given in our previous work \cite{ly}. Following
the same perturbative treatment as developed there. One can work out the
spectra of time dependent light scattering as given below.

\section{Spectrum of the scattered light}

In order to calculate the spectrum of the scattered light we need to first
determine the initial conditions for Eqs. (\ref{eqs8i}) and (\ref{eqs8}) at $%
t=-\infty$. We assume that initially the ground state energy levels were
populated according to the Fermi-Dirac distribution (FDD) for
non-interacting atoms in the harmonic well \cite{fddoo}. We also neglect the
much weaker $p$-wave collision interactions as they are energetically
suppressed in the temperature range of interests here.

The mean number of atoms in the $\vec n$-th level at $t=-\infty$ is
therefore
\begin{equation}  \label{bed}
N_{\vec n}=\langle g_{\vec n}^{\dag}g_{\vec n}\rangle ={\sf z} e^{-\beta
\omega_{\vec n}}/(1+{\sf z} e^{-\beta \omega_{\vec n}}),
\end{equation}
where $\beta=1/k_BT$, and ${\sf z}$ is the fugacity. The relation $\sum_{%
\vec n} N_{\vec n}=N$ determines ${\sf z}$ as a function of
$\beta$, $\omega_t$ and $N$.

We can now calculate the spectrum of scattered photons by using the
Hamiltonian (\ref{H}). We derive the Heisenberg equation for the photon
annihilation operator,
\begin{eqnarray}
{\dot a}_{\vec k\mu}&=&-ick a_{\vec k\mu}-i \varrho(k) \sum_{\vec n,\vec m%
}g_{\vec n}^{\dag}{\vec e}_{\vec m} \cdot{\vec\epsilon}_{\vec k\mu} \eta_{%
\vec n\vec m}({\vec k}),  \label{heisa}
\end{eqnarray}
and its Hermitian conjugate for $a_{\vec k\mu}^{\dag}$. These equations are
now solved perturbatively with respect to the atom-photon field coupling $%
\varrho(k)$. The formal solution of Eqs. (\ref{heisa}) is
\begin{eqnarray}
{a}_{\vec k\mu}(t)&=&e^{-ickt} a_{\vec k\mu}(-\infty)-i \varrho(k) \sum_{%
\vec n,\vec m}\eta_{\vec n\vec m}({\vec k})  \nonumber \\
&&\times\int_{-\infty}^{t}dt^{\prime}e^{-ick(t-t^{\prime})}{\vec\epsilon}_{%
\vec k\mu}\cdot{\vec e}_{\vec m}(t^{\prime}) g_{\vec n}^{\dag}(t^{\prime}).
\label{heisa2}
\end{eqnarray}
The perturbative solution is then obtained similar to the case of bosons
\cite{ly} but with care to take into account the anti-commutation relations
between fermionic atomic operators.

The total spectrum of scattered photons is defined as
\begin{equation}
C({\vec k},\mu)=\lim_{t\to \infty}\big\langle a^{\dag}_{{\vec k}\mu}(t) a_{{%
\vec k}\mu}(t)\big\rangle,
\end{equation}
and can be divided into coherent and incoherent parts,
\begin{equation}
C({\vec k},\mu)=C_{{\rm coh}}({\vec k},\mu)+C_{{\rm in}}({\vec k},\mu).
\label{coh1}
\end{equation}
The coherent part from $\langle a_{{\vec k}\mu}(t)\rangle$, as in the single
atom case, is proportional to the square modulus of the Fourier transform of
the mean atomic polarization. The incoherent part, however, is due to the
quantum fluctuations of the atomic polarization. Although in usual
experiments, only the total spectrum can be measured, the division into
coherent and incoherent parts is meaningful since they have significantly
different angular characteristics, as we show latter. The total spectrum is
\begin{eqnarray}  \label{spec1}
C({\vec k},\mu)&=& |\varrho(k)|^2|{\vec\epsilon}_L\cdot{\vec\epsilon}_{\vec k%
\mu}|^2  \nonumber \\
&\times& \sum_{\vec n_1,\vec m_1,\vec n_2,\vec m_2} \big[\eta_{\vec n_1\vec m%
_1}({\vec k})\big]^{*} \eta_{\vec n_2\vec m_2}({\vec k})  \nonumber \\
&\times& \sum_{\vec n_1^{\prime},\vec n_2^{\prime}} \eta_{\vec n_1^{\prime}%
\vec m_1}({\vec k}_L) \big[\eta_{\vec n_2^{\prime}\vec m_2}({\vec k}_L)\big]%
^{*}  \nonumber \\
&\times& \int_{-\infty}^{\infty}dt_1\int_{-\infty}^{\infty}dt_2
e^{-i(ck-\omega_L)t_1}e^{i(ck-\omega_L)t_2}  \nonumber \\
&\times& \big\langle {\vec f}_{\vec n_1^{\prime}}^{\,\dag}(t_1)\cdot{\vec%
\epsilon}_{L} g_{\vec n_1}(t_1)g_{\vec n_2}^{\dag}(t_2){\vec f}_{\vec n%
_2^{\prime}}(t_2) \cdot{\vec\epsilon}_{L}\big\rangle.
\end{eqnarray}
In the perturbative limit, we insert now the solutions of Eqs.
(\ref{eqs8i}) and (\ref{eqs8}) into Eq. (\ref{spec1}). Here, the
perturbation parameter is the coupling constant, $\rho(k)$, of
interactions leading to spontaneous emission. As the laser pulse
is assumed intense, its corresponding interaction is much
stronger, while the scattered light is weak so that it needs to be
treated quantum mechanically. The perturbation approach consists
of using zeroth order solutions for atomic operators in the
Heisenberg picture to find the first order quantum correction to
the photon operator. This method has also been applied in previous
studies of scattering problems in bosonic systems\cite{ly}.

At $t=-\infty$, the Heisenberg picture coincides with the
Schr\"odinger picture, so that we omit in the following the
explicit time dependence of the operators at $t=-\infty$. Since
initially all atoms are in the ground electronic state, we obtain
\begin{eqnarray}  \label{sol1}
&\ &\big\langle {\vec f}_{\vec n_1^{\prime}}^{\,\dag}(t_1)\cdot{\vec\epsilon}%
_{L} g_{\vec n_1}(t_1)g_{\vec n_2}^{\dag}(t_2){\vec f}_{\vec n%
_2^{\prime}}(t_2) \cdot{\vec\epsilon}_{L}\big\rangle  \nonumber \\
&=&\langle g_{\vec n_1^{\prime}}^{\dag}g_{\vec n_1} g_{\vec n_2}^{\dag}g_{%
\vec n_2^{\prime}}\rangle  \nonumber \\
&\times& \sin\big[A(t_1)\big]\cos\big[A(t_1)\big]\sin\big[A(t_2)\big] \cos%
\big[A(t_2)\big]  \nonumber \\
&+&\langle g_{\vec n_1^{\prime}}^{\dag} {\vec\epsilon}_L\cdot{\vec f}_{\vec n%
_1} {\vec\epsilon}_L\cdot{\vec f}_{\vec n_2}^{\dag} g_{\vec n%
_2^{\prime}}\rangle\sin^2\big[A(t_1)\big]\sin^2\big[A(t_2)\big],
\end{eqnarray}
where the expectation values $\langle\cdots\rangle$ with respect to the FDD
remains to be evaluated. Thus the only difference between the bosons and
fermions so far is the evaluation of the expectation values according to the
relevant statistics and initial conditions. The formal expressions are the
same.

The single atom spectrum can also be written as a sum of coherent
and incoherent parts \cite{ly},
\begin{equation}
S(\varpi) = S_{{\rm coh}}(\varpi) + S_{{\rm in}}(\varpi),
\end{equation}
with $\varpi=\delta\omega/\gamma_L=(ck-\omega_L)/\gamma_L$, and
for a hyperbolic secant pulse $1/\cosh(\gamma_Lt)$ of area $2\pi$,
\begin{eqnarray}
S_{{\rm
coh}}(\varpi)&=&|\rho(k_0)|^2(\vec{\epsilon}_{\vec{k}\mu}\cdot{\vec
\epsilon}_L)^2\pi \varpi^2/\cosh^2(\pi \varpi/2) , \\ S_{{\rm
in}}(\varpi)&=&|\rho(k_0)|^2(\vec{\epsilon}_{\vec{k}\mu}\cdot{\vec
\epsilon}_L)^2\pi \varpi^2/\sinh^2(\pi \varpi/2) . \nonumber
\end{eqnarray}
Since the results are only weakly dependent on a particular pulse
shape we shall present results for the hyperbolic secant pulse
only.

Using the relationship
\begin{equation}
\sum_{\vec m}\big[\eta_{\vec n^{\prime}\vec m}({\vec k})\big]^{*} \eta_{\vec %
n\vec m}({\vec k}_L)=\eta_{\vec n\vec n^{\prime}} (\vec k_L-\vec k),
\label{sumr2}
\end{equation}
we obtain
\begin{eqnarray}
C({\vec k},\mu)&=&\sum_{\vec n_1,\vec n_2,\vec n_1^{\prime},\vec
n_2^{\prime}} \big[\eta_{\vec n_1\vec n_1^{\prime}} ({\vec
k}-{\vec k}_L)\big]^{*} \eta_{\vec n_2 \vec n_2^{\prime}}({\vec
k}-{\vec k}_L)  \nonumber \\ &\times& \left[\langle g_{\vec
n_1^{\prime}}^{\dag}g_{ \vec n_1} g_{\vec n_2}^{\dag}g_{\vec
n_2^{\prime}}\rangle S_{{\rm coh} }(\varpi)\right.  \nonumber \\
&+&\left.\langle g_{\vec
n_1^{\prime}}^{\dag}{\vec\epsilon}_L\cdot{\vec f}_{ \vec n_1}
{\vec\epsilon}_L\cdot{\vec f}_{\vec n_2}^{\dag} g_{\vec n
_2^{\prime}}\rangle S_{{\rm in}}(\varpi)\right].  \label{sol2}
\end{eqnarray}
Making use of the properties of the FDD we find
\begin{eqnarray}
\langle g_{\vec n_1^{\prime}}^{\dag}g_{\vec n_1} g_{\vec n_2}^{\dag}g_{\vec n%
_2^{\prime}}\rangle &=&\delta_{\vec n_1^{\prime}\vec n_1} N_{\vec n_1}
\delta_{\vec n_2^{\prime}\vec n_2} N_{\vec n_2}  \nonumber \\
&+&\delta_{\vec n_1\vec n_2} \delta_{\vec n_1^{\prime}\vec n_2^{\prime}}N_{%
\vec n_1^{\prime}} \big(1- N_{\vec n_1}\big),  \label{sol3}
\end{eqnarray}
and
\begin{equation}
\langle g_{\vec n_1^{\prime}}^{\dag}{\vec\epsilon}_L\cdot{\vec f}_{\vec n_1}
{\vec\epsilon}_L\cdot{\vec f}_{\vec n_2}^{\dag} g_{\vec n_2^{\prime}}\rangle
=\delta_{\vec n_1\vec n_2}\delta_{\vec n_1^{\prime}\vec n_2^{\prime}} N_{%
\vec n_1^{\prime}}.  \label{sol4}
\end{equation}
The difference between bosons and fermions appears in the last term of Eq. (%
\ref{sol3}); $1-N$ for fermions and $1+N$ for bosons
\cite{ly}. In driving the above expression we used the fact that
$N_{\vec{n}}^2 = \langle (g^\dag_{\vec n} g_{\vec %
n})^2\rangle = N_{\vec{n}}$ for fermions since the occupation of
any given level is either $0$ or $1$. This implies that the
quantum dispersion of fermions is given by $\delta N_{\vec{n}}^2 =
N_{\vec{n}}(1-N_{\vec{n}})$.

Inserting the above expressions in Eq.~(\ref{sol2}), and performing tedious,
but elementary calculations we finally obtain analytic expressions for the
spectra. In particular, the coherent part is
\begin{eqnarray}
C_{{\rm coh}}({\vec k},\mu)=S_{{\rm coh}}(\varpi) \Big| \sum_{\vec n} N_{%
\vec n}\, \eta_{\vec n\vec n}({\vec k}-{\vec k}_L)\Big|^2,  \label{spcoh}
\end{eqnarray}
identical in form as the bosonic system, but now with $N_{\vec n}$
representing the mean occupation number for a fermionic system.

The incoherent part of the spectrum is
\begin{eqnarray}
C_{{\rm in}}({\vec k},\mu) &=& S_{{\rm coh}}(\varpi)\sum_{\vec n}
\sum_{\vec n^{\prime}} N_{ \vec n}(1-N_{\vec n^{\prime}})
\big|\eta_{\vec n\vec n^{\prime}}({\vec k}-{\vec k}_L)\big|^2
\nonumber \\ &+& NS_{{\rm in}}(\varpi). \label{spin}
\end{eqnarray}
Again the difference appears in the $1-N$ term on the first line of Eq. (%
\ref{spin}) compared with $1+N$ in the case of bosons \cite{ly}. We note
that the incoherent spectrum (\ref{spin}) consists of three parts coming
from: i) quantum dispersion of the occupation numbers $\delta N_{\vec n}^2=
\langle (g_{\vec n}^{\dag}g_{\vec n})^2\rangle- \langle g_{\vec n}^{\dag}g_{%
\vec n}\rangle^2$, ii) processes of creation of the $n$-th wave packet
accompanied by annihilation of the ${\vec n}^{\prime}$-th one for ${\vec n}%
\ne {\vec n}^{\prime}$, and iii) the single atom incoherent
spectrum. Obviously, both coherent and incoherent spectra reflect
quantum statistical properties of atoms since they depend on
$N_{\vec n}$'s, which are described by the Fermi-Dirac
distribution for our system of fermions. The first term in Eq.
(\ref{spin}) for the incoherent spectrum depend explicitly on the
statistical properties of atoms.

The total number of emitted photons can be obtained by integrating the
spectrum,
\begin{equation}
N_{{\rm tot}}=\sum_{\mu}\int d{\vec k}\;C({\vec k},\mu),  \label{litot}
\end{equation}
which could also be divided into coherent and incoherent parts. By fixing
the direction of $\vec k$ and integrating over the azimuthal angle $\varphi$
one can also define an angular distribution of photons $dN_{{\rm tot}%
}(\theta)$ [and, correspondingly $dN_{{\rm coh}}(\theta)$ and $dN_{{\rm in}%
}(\theta)$]
\begin{eqnarray}
dN_{{\rm tot}}(\theta) &=&dN_{{\rm coh}}(\theta) +dN_{{\rm in}}(\theta)
\nonumber \\
&=&\sin(\theta)d\theta\sum_{\mu}\int_{0}^{2\pi}d\varphi \int k^2dk\;C({\vec k%
},\mu),  \label{dN_tot}
\end{eqnarray}
where $\theta$ is the angle between $\vec k$ and $\vec k_L$. We can also
choose to define an integrated spectrum by fixing $|\vec k|$, and
integrating the spectrum over the full solid angle.

\section{Numerical calculation of the spectrum}

The analytical expressions for both the coherent and incoherent spectra as
obtained in Eqs. (\ref{spcoh}) and (\ref{spin}) are not directly applicable
to straight numerical summations as they involve triple and six-fold sums
respectively. In the previously work for trapped bosons, the analogous
expressions were in terms of a power series of the fugacity ${\sf {z}}$. A
subsequent resummation of an auxiliary series provided a fast convergent
numerical approach. However, this same technique is only applicable when $%
{\sf z}< 1$ corresponding to a high temperature limit for the Fermi gas. We
therefore needed a new method to calculate the spectrum at low temperatures
when ${\sf z}\gg 1$, which is exactly the region where the interesting
effects of quantum statistics becomes important.

The coherent spectrum as expressed earlier in Eq. (\ref{spcoh}) consists of
a triple sum. Similar to the case for bosons \cite{ly}, when ${\sf z}< 1$,
we can write the spectrum as a power expansion of ${\sf {z}}$
\begin{eqnarray}
&&C_{{\rm coh}}({\vec k},\mu)=S_{{\rm coh}}(\varpi)
\left|\sum_{l=1}^{\infty}(-1)^{l-1}\frac{{\sf z}^l} {(1-e^{-l\beta\omega_{%
{\rm t}}})^3}\right.  \nonumber \\
&&\left.\times\exp\big[-\frac{1}{2} ({\vec k}-{\vec k}_L)^2a^2\coth(l\beta%
\omega_{{\rm t}}/2)\big]\right|^2.  \label{prop}
\end{eqnarray}
Thus we have transformed a triple sum into a single sum. Also for very high
temperatures when ${\sf z}\ll 1$, the above single sum converges quickly. On
the other hand, the quantum degenerate low temperature limit for fermions
corresponds to ${\sf z}\rightarrow\infty$ when zero temperature is
approached. We thus use an alternative method where the original expression
for the spectrum, Eq. (\ref{spcoh}), is written in terms of the Laguerre and
the generalized Laguerre polynomials ${\cal L}_n(.)$ and ${\cal L}_n^m(.)$.
The properties of these polynomials are then used to reduce the triple sum
into a single sum over the generalized Laguerre polynomials. The new form is
then (see Appendix A for details),
\begin{eqnarray}
C_{{\rm coh}}({\vec k},\mu)&=&S_{{\rm coh}}(\varpi) e^{-({\vec k}-{\vec k}%
_L)^2a^2}  \nonumber \\
&\times & \left|\sum_{n=0}^{\infty}P(n){\cal L}_n^{2} [({\vec k}-{\vec k}%
_L)^2a^2] \right|^2,  \label{coh_lag}
\end{eqnarray}
where
\begin{equation}
P(n)=\frac{{\sf z}e^{-\beta\omega_t n}}{1+{\sf z}e^{-\beta\omega_t n}}
\end{equation}
is the mean occupation number of fermions in any of the degenerate energy
states with principle quantum number $n$.

By using this new expression we achieve enormous savings in computational
time over the initial triple sum. However the added complexity of generating
the generalized Laguerre polynomials implies that this method is not as fast
as the power expansion method which was only valid for ${\sf z}<1$ \cite{ly}.

The evaluation of the incoherent spectrum is a more difficult numerical
problem as we see that the second term of Eq. (\ref{spin}) involves a
six-sum. This is due to the fact that incoherent photon emissions
corresponds to different final and initial motional energy levels (unlike
the coherent case where the two are the same). For the case of ${\sf {z}<1}$
we can again use the power expansion approach to write the incoherent
spectrum as
\begin{eqnarray}
C_{{\rm in}}(\vec k,\mu)&=&NS_{{\rm coh}}(\varpi)+NS_{{\rm in}}(\varpi)
\nonumber \\
&-&S_{{\rm coh}}(\varpi) \sum_{l_1,l_2=1}^{\infty}\frac{{\sf (-z)}^{l_1+l_2}%
}{[1-e^{-(l_1+l_2) \beta\omega_{{\rm t}}}]^3}  \nonumber \\
&\times&\exp\big[-(\vec k-\vec k_L)^2a^2 f(\beta,l_1,l_2)\big],
\label{inco1}
\end{eqnarray}
with
\begin{eqnarray}
f(\beta,l_1,l_2)=\frac{(1-e^{-l_1\beta\omega_{{\rm t}}})
(1-e^{-l_2\beta\omega_{{\rm t}}})} {1-e^{-(l_1+l_2)\beta\omega_{{\rm t}}}}.
\end{eqnarray}
We note that the minus `-' sign in front of the third term in
Eq.~(\ref {inco1}) is due to quantum statistics. In the bosonic
case considered earlier \cite{ly} a plus `+' sign was involved.
This approach helped to reduce the six-sum into a double sum.
Again this limit is only applicable for high temperatures (when
average energy per atom is much greater than the Fermi
temperature). In the general case, some alternative
simplifications are possible. We rewrite Eq. (\ref{spin}) as
\begin{eqnarray}
C_{{\rm in}}({\vec k},\mu) &=& NS_{{\rm coh}}(\varpi)+NS_{{\rm in}}(\varpi)
\nonumber \\
&-& S_{{\rm coh}}(\varpi)\sum_{\vec{n}}\sum_{\vec{n}^\prime} N_{\vec{n}}N_{%
\vec{n}^\prime} \big|\eta_{\vec n\vec n^{\prime}}({\vec k}-{\vec k}_L)\big|^2
\nonumber \\
&=& NS_{{\rm coh}}(\varpi)+NS_{{\rm in}}(\varpi)  \nonumber \\
&-&S_{{\rm coh}}(\varpi) \sum_{n_x,n_y,n_z}\sum_{m_x,m_y,m_z}  \nonumber \\
&\times& P_{{\rm inc}}(n_x+n_y+n_z,m_x+m_y+m_z)  \nonumber \\
&\times& \big| {\cal I}_{n_x,m_x}{\cal I}_{n_y,m_y}{\cal I}_{n_z,m_z} \big|%
^2,  \label{spin2}
\end{eqnarray}
with
\begin{equation}
P_{{\rm inc}}(n,m)= \frac{{\sf z}e^{-\beta\omega_t n}\,{\sf z}%
e^{-\beta\omega_t m}} {(1+{\sf z}e^{-\beta\omega_t n})(1+{\sf z}%
e^{-\beta\omega_t m})},
\end{equation}
and
\begin{equation}
{\cal I}_{n_j,m_j}=\langle n_j|e^{-i\delta k_j\cdot R_j} |m_j\rangle.
\end{equation}
Where we denote the $j$-th component of $\delta \vec k= \vec k-\vec k_L$ by $%
\delta k_j$. The number of sums can be reduced by exploiting the symmetry of
the scattering geometry. Because the scattering is symmetric about the axis
along the incoming laser direction $\vec k_L$. We can set one of the $\delta%
\vec k$ components in the plane perpendicular to the laser axis to be zero,
i.e. choosing the laser to be aligned with the $z$-axis we can set $k_y=0$.
The incoherent spectrum with the aid of this symmetry and the identity Eq. (%
\ref{identity}) is then
\begin{eqnarray}
C_{{\rm in}}({\vec k},\mu) &=& NS_{{\rm coh}}(\varpi)+NS_{{\rm in}}(\varpi)
\nonumber \\
&-&S_{{\rm coh}}(\varpi)e^{-|\delta k|^2 a^2}  \nonumber \\
&\times&\sum_{n_x,n_z}\sum_{m_x,m_z} {\cal P}(n_x+n_z,m_x+m_z)  \nonumber \\
&\times& \big( \frac{m_x!m_z!}{n_x!n_z!}\big) (\delta k_x^2
a^2)^{n_x-m_x}(\delta k_z^2 a^2)^{n_z-m_z}  \nonumber \\
&\times& \big|{\cal L}_{n_x}^{n_x-m_x}(\delta k_x^2 a^2) {\cal L}%
_{n_z}^{n_z-m_z}(\delta k_z^2 a^2) \big|^2  \label{spin3}
\end{eqnarray}
with
\begin{equation}
{\cal P}(n,m)=\sum_{y}P_{{\rm inc}}(n+y,m+y).
\end{equation}
This finally reduces the six-sum to a four-sum.

\section{Discussion of the spectrum}

In this section we discuss three methods for analyzing the spectrum. 1) we
look at the part of the spectrum that reveals quantum statistics. This
represents all the collective effects on the spectrum. 2) we integrate over
either the frequency (angle) to obtain an angular (frequency) spectrum
respectively. 3) the total number of scattered photons is calculated to
illustrate the overall scattering behavior as a function of the temperature
and hence the degeneracy of trapped fermions.

\subsection{Form functions}

A useful part of the spectrum is the component that describes the quantum
statistics as opposed to the single atom component of the spectrum. We call
this component the ``form function" as it is indeed related to the Fourier
transform of the average density profile of the degenerate gas. Temperature
dependent behavior of the scattering spectrum, the transition from classical
statistics at high temperatures to the Fermi-Dirac one at low temperatures,
will manifest itself in the form function.

\begin{figure}[t]
\centerline{\epsfig{file=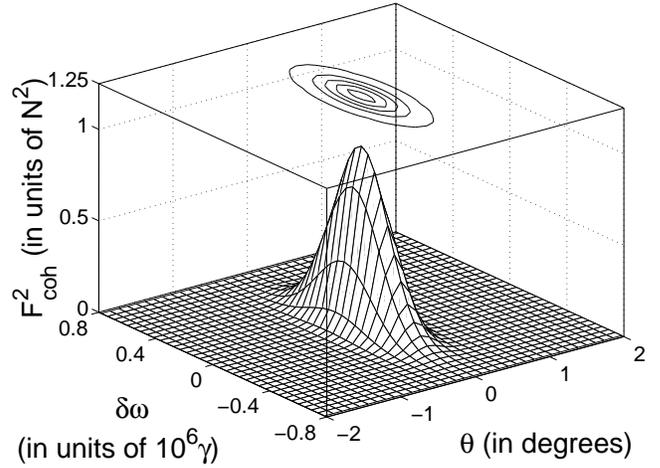,width=8.5cm}\\[12pt]}
\caption{The coherent form function for one million trapped
fermionic atoms at a temperature of $k_BT/E_F=1.36$.}
\label{fig01}
\end{figure}

\begin{figure}[t]
\centerline{\epsfig{file=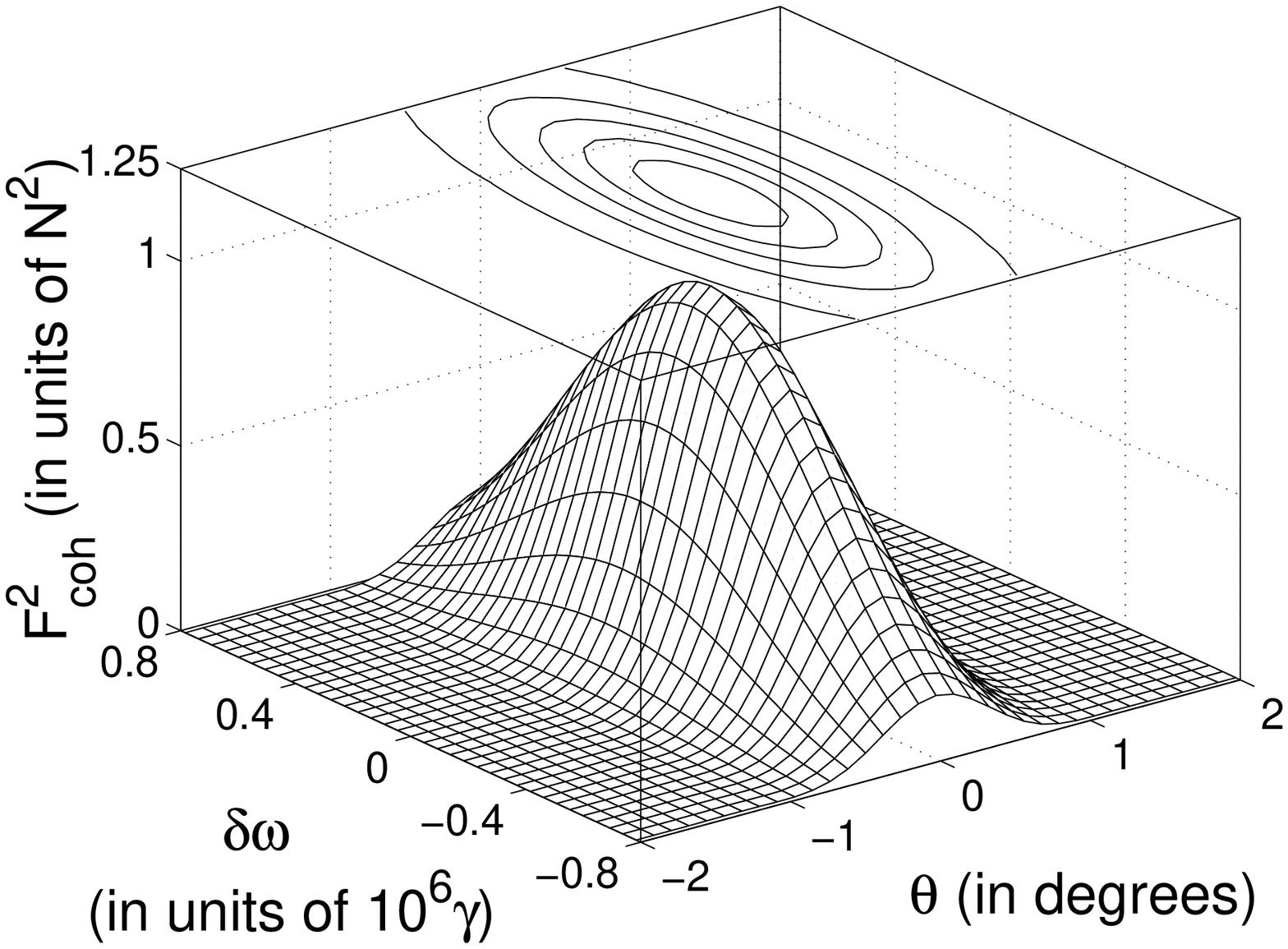,width=8.5cm}\\[12pt]}
\caption{The same as Fig.~1 but at a temperature of
$k_BT/E_F=0.0016$.} \label{fig03}
\end{figure}

In Eq. (\ref{prop}), the quantum statistical component is the expression to
the right of the single atom term $S_{{\rm coh}}(\varpi)$. Thus for ${\sf z}%
<1$ we have the following form function
\begin{eqnarray}
&&{\cal F}_{{\rm coh}}^{2}(\delta\omega,\theta)\nonumber\\
&=&\left|\sum_{m=1}^{\infty}(-1)^{m-1}\frac{{\sf z}^m} {(1-e^{-m\beta\omega_{%
{\rm t}}})^3}\right.  \nonumber \\
&\times&\left.\exp\big[-\frac{1}{2} ({\vec k}-{\vec k}_L)^2a^2\coth(m\beta%
\omega_{{\rm t}}/2)\big]\right|^2.  \label{Fcoh}
\end{eqnarray}
For atoms obeying Maxwell-Boltzman distribution, the above
summation can be analytically evaluated to give
\begin{eqnarray}
{\cal F}_{{\rm coh}}^{2}(\delta\omega,\theta) = N^2
\exp{\left[-({\vec k}-{\vec k}_L)^2a^2\coth(\beta \omega_{{\rm
t}}/2)\right]}.
\end{eqnarray}
Thus, at high temperatures coherent spectrum of scattered light
from fermionic atoms should coincide with the classical results.
For the general case of arbitrary temperature we can rewrite Eq.
(\ref{coh_lag}) as
\begin{eqnarray}
{\cal F}_{{\rm coh}}^{2}(\delta\omega,\theta)&=& e^{-({\vec
k}-{\vec k} _L)^2a^2}  \nonumber \\ &\times&
\left|\sum_{n=0}^{\infty}P(n){\cal L}_n^{2} [({\vec k}-{\vec k}
_L)^2a^2] \right|^2.  \label{Fcoh2}
\end{eqnarray}
Figures~\ref{fig01}-\ref{fig03} represent the coherent form
functions for temperatures of $k_BT/E_{{\rm F}}=1.36$, and $
0.0016$ with a system consisting of one million atoms. Results are
presented as rescaled by $N^2$, since ${\cal F}^{2}_{\rm
coh}(0,0)=N^2$, independent of temperature. In all numerical
simulations we have set the dimensionless parameter $k_L a=12.5$.
For a resonant transition wavelength of $\sim 800$ (nm), our
choice corresponds to magnetic trapping frequencies of about
$2\pi\times 300$ (Hz), $2\pi\times 50$ (Hz), and $2\pi\times 23$
(Hz) for $^6$Li, $^{40}$K, and $^{86}$Rb respectively. The ground
state trap size is $\sim 1.6$ ($\mu$m). We see that the form
function becomes broader as the temperature drops. At high
temperatures phase matching effects become dominant where
destructive interference attenuates the scattering except for a
narrow cone in the forward direction. As $k_BT$ drops below the
Fermi energy $E_{{\rm F}}$, phase matching becomes less important
and the role of the quantum statistics of the gas becomes
increasingly significant. As a rule of thumb, the effects
of quantum statistics comes into play when $k_BT$ is less than one-half of $%
E_{{\rm F}}$, consistent with the effects seen in evaporative
cooling \cite {geist} and recent experimental studies \cite{Jinn}.
We note that once the quantum statistic effects become
significant, lowering the temperature does little to the shape of
the form function as the lowest energy levels become close to
being ``stacked", and start to block further filling due to the
Pauli exclusion principle. Any additional cooling will therefore
only result in filling of the abundant higher energy levels to
enlarge the Fermi sea.

\begin{figure}[t]
\centerline{\epsfig{file=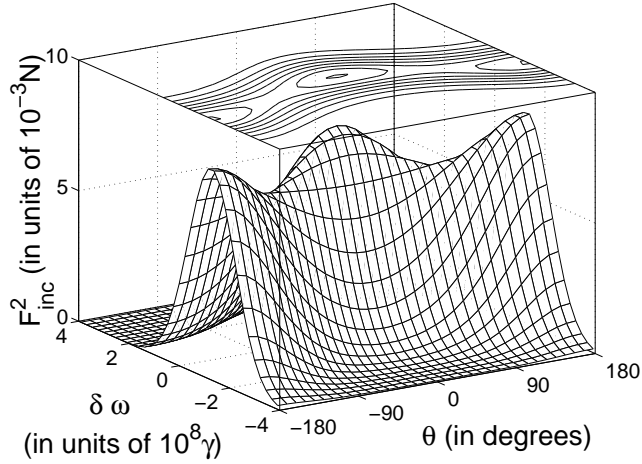,width=8.5cm}\\[12pt]}
\caption{The incoherent form function for one million trapped
fermionic atoms at a temperature of $k_BT/E_F=1.36$. Note that we
have scaled the form function by $N$ instead of $N^2$.}
\label{fig04}
\end{figure}

\begin{figure}[b]
\centerline{\epsfig{file=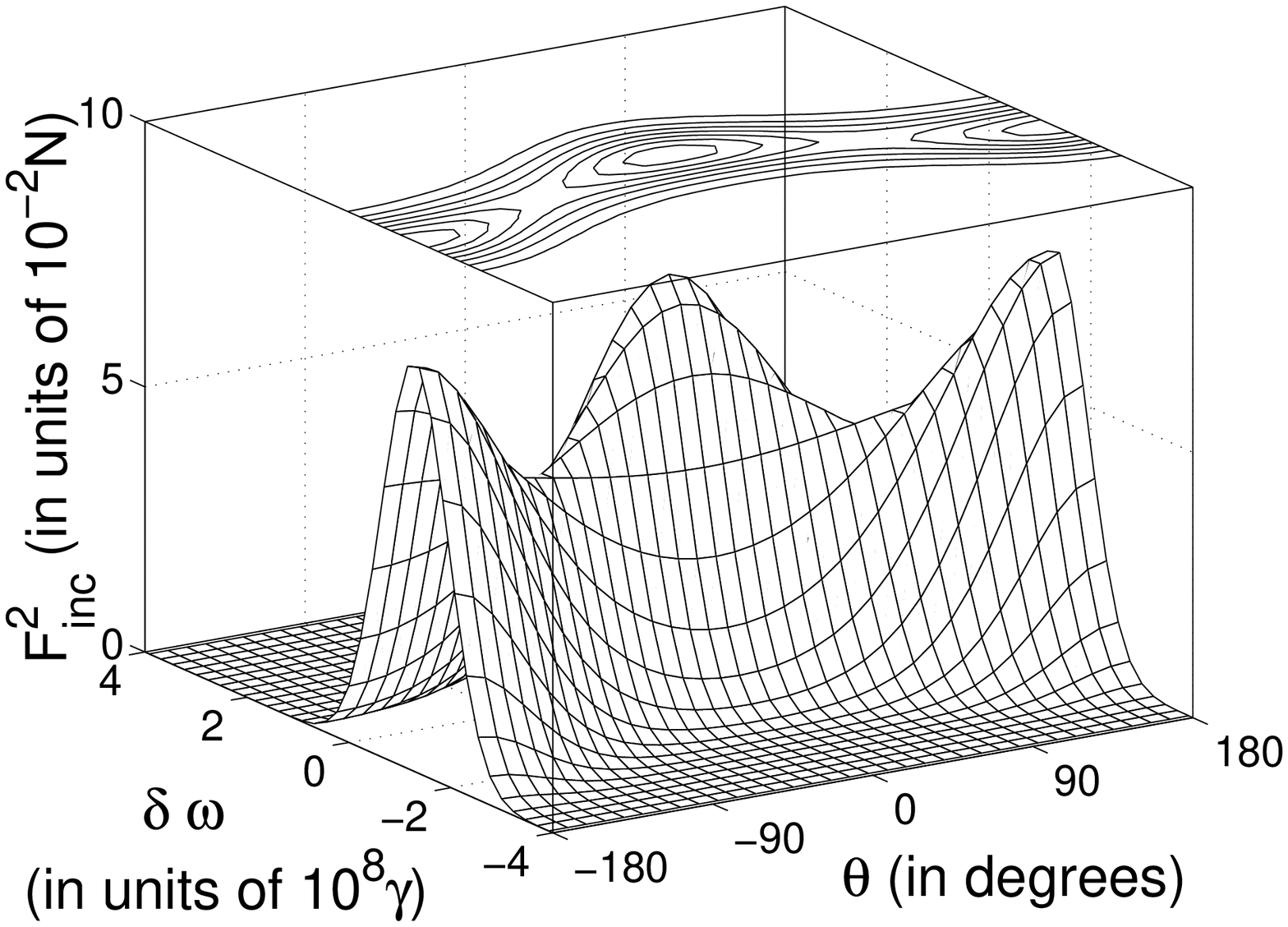,width=8.5cm}\\[12pt]}
\caption{The same as Fig.~4 but at a temperature of
$k_BT/E_F=0.6$.} \label{fig05}
\end{figure}

The form function for the incoherent spectrum when ${\sf z}<1$ is retrieved
from Eq.~(\ref{inco1})
\begin{eqnarray}
{\cal F}_{{\rm inc}}^{2}(\delta\omega,\theta )
&=&\sum_{l_{1},l_{2}=1}^{\infty }\frac{{\sf (-z)}^{l_{1}+l_{2}}}{%
[1-e^{-(l_{1}+l_{2})\beta \omega _{{\rm t}}}]^{3}}  \nonumber \\
&\times &\exp \big[-(\vec{k}-\vec{k}_{L})^{2}a^{2}f(\beta ,l_{1},l_{2})\big].
\label{Finc}
\end{eqnarray}
The corresponding form function for the general case from Eq.~(\ref{spin3})
would be
\begin{eqnarray}
{\cal F}_{{\rm inc}}^{2}(\delta\omega,\theta )&=&e^{-|\delta
k|^{2}a^{2}}  \nonumber \\
&\times &\sum_{n_{x},n_{z}}\sum_{m_{x},m_{z}}{\cal P}%
(n_{x}+n_{z},m_{x}+m_{z})  \nonumber \\
&\times &\big( \frac{m_{x}!m_{z}!}{n_{x}!n_{z}!}\big) (\delta
k_{x}^{2}a^{2})^{n_{x}-m_{x}}(\delta k_{z}^{2}a^{2})^{n_{z}-m_{z}}  \nonumber
\\
&\times &\big| {\cal L}_{n_{x}}^{n_{x}-m_{x}}(\delta k_{x}^{2}a^{2}){\cal L}%
_{n_{z}}^{n_{z}-m_{z}}(\delta k_{z}^{2}a^{2})\big|^{2}.  \label{Finc2}
\end{eqnarray}
Figures~\ref{fig04}-\ref{fig05} display the incoherent form
functions for two different temperatures $k_{B}T/E_{{\rm F}}=1.36$
and $0.6$ for one million atoms. The form function at the cooler
temperature (Fig.~\ref{fig05}) is narrower
along the frequency axis and have a higher sharper peak than that of Fig.~%
\ref{fig04}. At the high temperature limit we can estimate the
height of the peak in Fig.~\ref{fig04} by using Maxwell-Boltzmann
statistics to find that ${\cal F}^{2}_{\rm in}(0,0)/N=N/(2kT)^3$.
Putting in the numbers for Fig.~\ref{fig04}; $N=10^6$ and
$1/(kT)=4.036\times 10^{-3}$ we obtain ${\cal F}^{2}_{\rm
in}(0,0)/N=8.2\times 10^{-3}$ which is consistent with the
numerical peak of $8.0\times 10^{-3}$. We also know that in
opposing limit of zero temperature the form function has a value
of $N$ at the peak. At frequency regions far away from the central
peak location, the form function is mainly governed by the rapidly
decaying exponential factor $\exp[{-({\vec k}-{\vec k} _L)^2a^2}]$
in Eqs. (\ref{Fcoh2}) and (\ref{Finc2}). This is so because the
weighting factors ${\cal P}$ of the equilibrium atom number
occupations more than off-set the asymptotically growing
($x^{n}/n!$) behaviors of the Laguerre polynomials $L_{n}^{\alpha
}(x)$. Consequently, higher order Laguerre polynomials don't play
any significant role in determining the asymptotic character of
the form function, and the form function simply decays
exponentially which physically corresponds to suppressed
scatterings into too wide frequency separations. On the other
hand, angular behavior of form function shows periodical
character.

We see that the form functions are asymmetric along the frequency
domain but remains symmetric in angular one. This angular symmetry
is basically due to our previous assumption on the cylindrical
symmetry of the scattering, since the exponential factor
$\exp[{-({\vec k}-{\vec k} _L)^2a^2}]$ depends on angle $\theta$
through the even function of $\cos{\theta}$. The asymmetry along
the frequency direction is also consistent with the asymmetry of
exponential factor with respect to frequency $\omega=c|\vec{k}|$.

\subsection{Angular and frequency spectrum}

The angular and frequency spectra are of interest as they can be easily
observed in an experiment. We have already defined an angular spectrum of
photons $dN_{{\rm tot}}(\theta)$ previously by Eq.~(\ref{dN_tot}). Note that
the domains of the variables; the azimuthal angle $\varphi$, polar angle $%
\theta$ and $k$ the magnitude of $\vec{k}$ are: $\varphi\in [0,2\pi]$, $%
\theta\in[0,\pi]$ and $k\in [0,\infty]$. It is desirable to look at the
spectrum of the coherent and incoherent components of the spectrum
separately as their behavior are significantly different. We do not
explicitly write down the expressions for the coherent situation as it is
trivial; it involves simply an integration and sum over the polarizations of
the product of the single atom contribution $S_{{\rm coh}}$ and the coherent
form function ${\cal F}^2_{{\rm coh}}$. The incoherent situation is, on the
other hand, a little more complicated. Using the expressions for $C_{{\rm in}%
}$ from Eq.~(\ref{inco1}) and Eq.~(\ref{spin3}) we have the following
general expression
\begin{eqnarray}
dN_{{\rm in}}(\theta)&=&\pi(1+\cos^2\theta)\sin\theta d\theta
\nonumber \\
&\times&\big\{N\int_{0}^{\infty}\big [S_{{\rm coh}}(\varpi) +S_{{\rm in}%
}(\varpi)\big ]\,k^2 dk  \nonumber \\
&-&\int_{0}^{\infty}S_{{\rm coh}}(\varpi){\cal F}_{{\rm inc}%
}^{2}(\theta,k)\, k^2 dk\big\},  \label{spec_th}
\end{eqnarray}

To obtain the analogous expression for the angle integrated frequency
spectrum one simply integrate over $\theta$ instead of $k$ in Eq.~(\ref
{spec_th}).

\begin{figure}[t]
\centerline{\epsfig{file=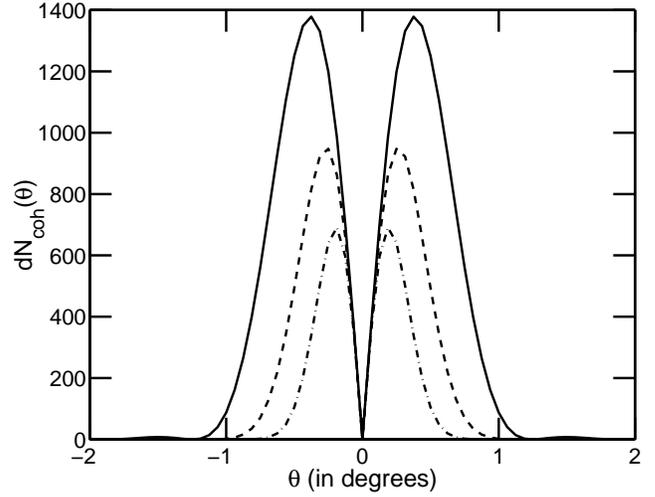,width=8.5cm}\\[12pt]}
\caption{The differential coherent scattering versus angle for ten
thousand trapped fermionic atoms. The solid, dashed, and
dash-dotted curves represent temperatures of $k_BT/E_{F}=0.001$,
$0.5$ and $1.0$ respectively.} \label{fig06}
\end{figure}

\begin{figure}[b]
\centerline{\epsfig{file=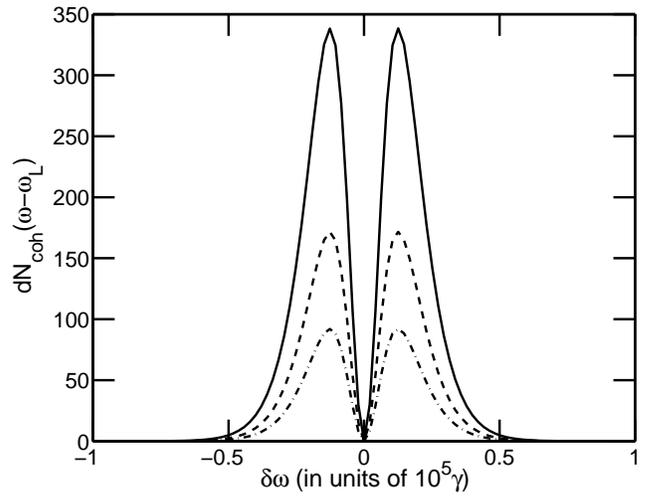,width=8.5cm}\\[12pt]}
\caption{The differential coherent scattering as a function of
frequency for ten thousand trapped fermionic atoms. Follows the
same curve format and temperatures as Fig.~\ref{fig06}.}
\label{fig07}
\end{figure}

Figures~\ref{fig06} and \ref{fig07} show coherent angular and
frequency spectra for one million atoms with a laser pulse width
of $\tau_L=10$ ps. As shown in Fig.~\ref{fig06}, the coherent
angular spectrum is narrow with a range from about $-2$ to $2$
degrees. The solid, dashed, and dash-dotted curves corresponds to
temperatures of $k_BT/E_{F}=0.001$, $0.5$, and $1.0$ respectively.
Curves for lower temperatures are broader and greater in magnitude
than the high temperature dash-dotted curve. Figure~\ref{fig07}
shows the frequency spectrum for the same three temperatures using
the same curve formats. The overall features for the coherent
spectra are the increase in magnitude and broadening of the
scattering as the temperature drops. As explained previously in
the form function section, the effects of phase matching becomes
diminished and quantum statistical effects becomes more dominant
as the temperature cools leading to an observable broadening and
an overall increased scattering.

\begin{figure}[t]
\centerline{\epsfig{file=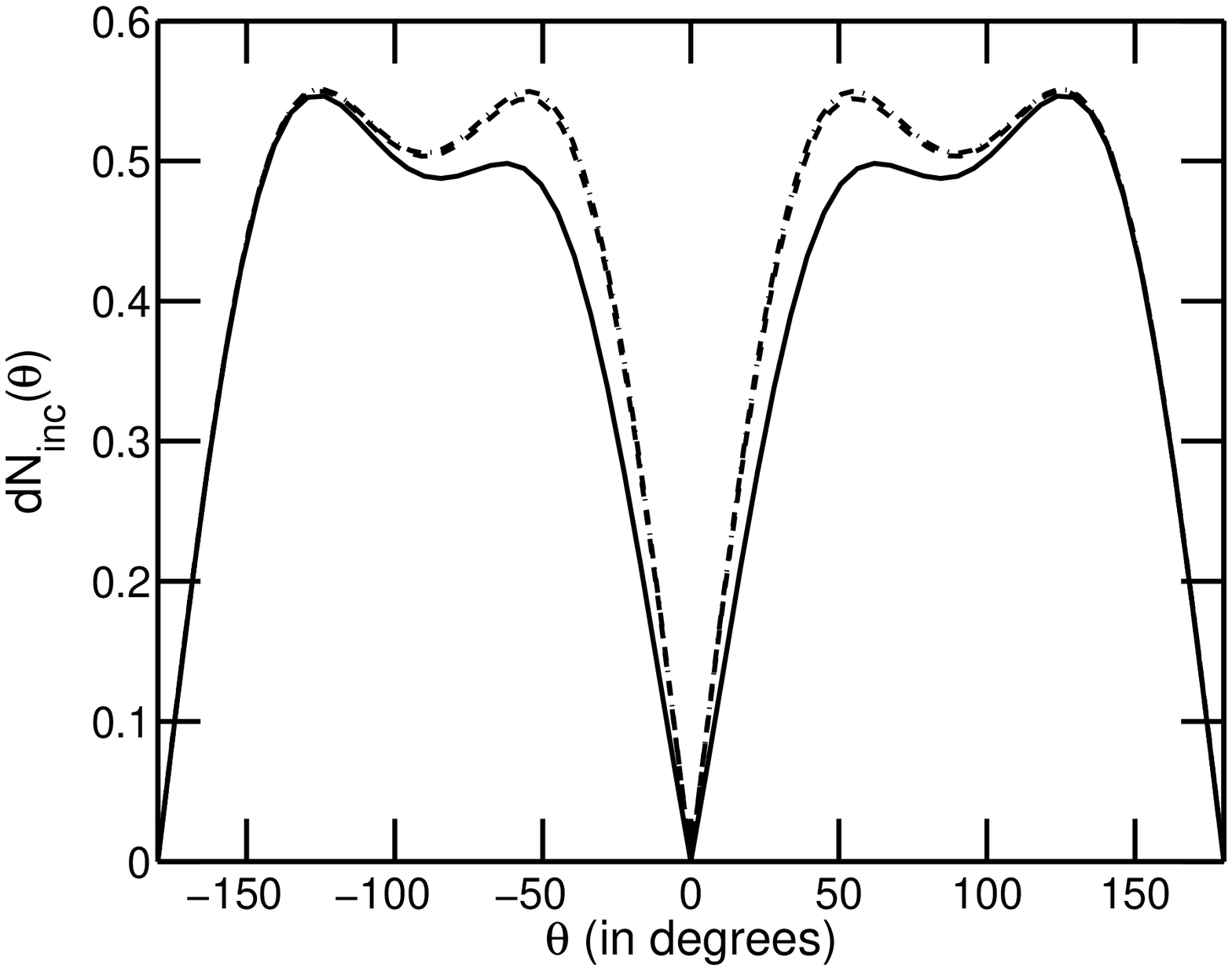,width=8.5cm}\\[12pt]}
\caption{The same as Fig.~\ref{fig06} but for differential
incoherent scattering.} \label{fig08}
\end{figure}

\begin{figure}[b]
\centerline{\epsfig{file=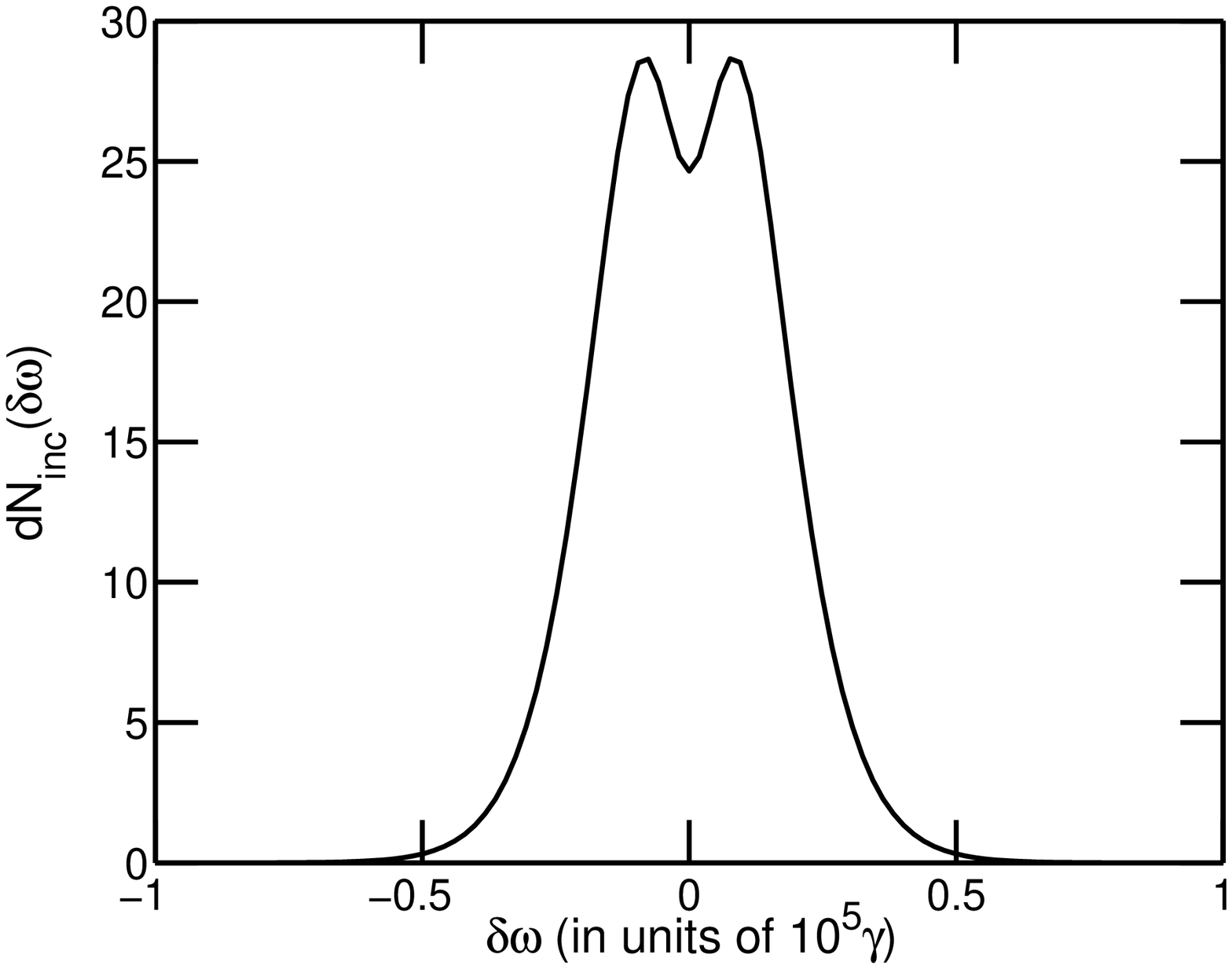,width=8.5cm}\\[12pt]}
\caption{The same as Fig.~\ref{fig07} but for differential
incoherent scattering.} \label{fig09}
\end{figure}

Finally Figs. \ref{fig08} and \ref{fig09} show  incoherent angular
and frequency spectra for one million atoms with a laser pulse
width of $ \tau_L=10$ ps. The incoherent angular spectrum is far
broader than the coherent one with a range from about $-180$ to
$180$ degrees, covering the complete polar range. We used the same
curve formats as before for the three temperatures. We see that
all three curves are almost on top of each other for angular
ranges $90<\theta<180$ and $-180<\theta<-90$, which corresponds to
back-scattering regime. The single atom spectra $S_{\rm coh},
S_{\rm in}$ are appreciable only within a frequency scale $\sim
10^5\gamma$, much shorter than the frequency scale $\sim
10^8\gamma$ for any significant change in form functions $F_{\rm
coh}^2$ and $F_{\rm in}^2$. Thus, frequency dependence of form
functions cannot play a role in determining the behavior of the
spectrum of the scattered photons. We approximate from
Eq.(\ref{spec_th}) that angular behavior of differential spectrum
follows $(1+\cos^{2}{\theta})|\sin{\theta}|[N-F^{2}_{\rm
in}(0,\theta)]$. Therefore, we conclude that the decrease in the
first peak is due to the increase of $F^2_{\rm in}$ at large
number of atoms and at low temperatures. The second peak is not
affected since within the frequency scale of single atom spectra,
the form function is already diminishingly small at those angles.
The minus sign in $N-F^{2}_{\rm in}(0,\theta)$ is due to
anti-commutation of fermionic atomic operators, thus is purely
quantum statistical. In Fig.~\ref{fig09} we see that again all
three curves follow one another closely. However they now possess
a dip at the resonance frequency. Because these spectrum consists
of incoherent processes, unlike the case of coherent scattering at
high temperatures, there is no phase matching effects. The dip in
the frequency spectrum is mathematically due to the requirement
that all the three temperature curves meet at $\omega-\omega_0$
equal zero. For this particular point the form functions and hence
the spectra does not depend on the temperature, reminiscent of
some kind of optical theorem \cite{ly}.

\subsection{Number of scattered photons}

The total number of scattered photons is also directly observable in an
experiment. By calculating the total number of scattered photons from a
simple pulse excitation we can study its dependence on the temperature. We
again separate these into coherent and incoherent components. The number of
scattered photons scales as $N^2$ for the coherent case and as $N$ for the
incoherent one. For larger numbers of atoms one expects that the scattering
will be dominated by the coherent scattering in the short pulse limit.
Similarly, for low atom numbers, the incoherent scattering will become
dominate and thus be observable when only few atoms are scattered.

The coherent scattering becomes more dominant at lower
temperatures for the system of one million atoms. Our preference
for a dominant coherent scattering is due to the more sensitive
nature of the coherent scattering to changes in temperature. This
is shown clearly in the following figure, Fig. \ref{fig11} showing
coherent and incoherent scattering as a function of the
temperature for one million atoms. The coherent results are
plotted as circles while squares denote the incoherent ones. For
comparison the triangle plots are calculated using the
Maxwell-Boltzmann distribution (MBD) for a classical gas. The
coherent curve displays temperature-dependent sensitivity across
the full range of temperatures both above and below $E_F$. The
incoherent curve, on the other hand, is flat across the entire
range of temperatures that we used. As is expected, the MBD and
FDD coherent curves are the same at high temperatures since
quantum statistical nature of the atom becomes less important as
$k_BT/E_F$ is greater than $0.5$. They start to deviate from each
other for $k_BT/E_F$ between $0$ and $0.5$. This becomes very
pronounced at very low temperatures when $k_BT/E_F$ is less than
one-tenth. The FDD curve (circles) displays a flattening off at
very low temperatures. The MBD curve, on the other hand, does not
share this feature, but it displays a rapid increase in scattering
at these low temperatures. Thus the flattening off is caused by
the fermionic nature of the atoms. Previous work on a bosonic
system displayed a dramatic increase scattering at these low
temperatures. This flattening originates from the inhibition of
spontaneous emission far below the Fermi temperatures where the
lower levels are ``stacked" so that spontaneous emission into
these levels are forbidden by the Pauli's exclusion principle
\cite{zoller}.

\section{Conclusions}

We have studied theoretically the scattering of intense short
laser pulses off a system of cold atoms. We have presented a
detailed theory of such processes. We have demonstrated that by
scattering pulses of area $2\pi K$ one may observe signatures of
fermionic degeneracy in a system of trapped atoms. In the regime
of validity of our theory, $2\pi K$ pulses leave the system of
trapped atoms practically unperturbed. At high temperatures when
the average energy per atom is many times the Fermi temperature
$T_F$ the coherent scattering is very weak and occurs in a very
narrow cone in the forward direction due to phase matching
effects, which is the same effect as for the case of bosons
\cite{ly}. As the temperature $k_BT$ becomes of order less than
one half of $E_F$, angular distributions of scattered coherent
photons broadens as the influence of the phase matching effects
are reduced. Incoherent scattering is insensitive, in both angular
and frequency spectra (hence total photons scattered) to the
temperature at small atom numbers $(\leq 10^5)$. For such small
number of atoms incoherent scattering dominates and scattering of
short laser pulses is weakly temperature dependent. On the other
hand, we find for larger number of atoms ($\ge 10^6$ atoms with
other parameters chosen as in this study), the angular incoherent
spectrum, as well as the total number of scattered photons,
exhibit qualitative effects depending on temperature.

The number of scattered photons increases as the system of atoms are cooled
but the rate of this increase tails off as the fermionic nature sets in to
suppress photon scattering events leading to atoms into already occupied
motional states. This temperature dependent property is compared with
results calculated for atoms obeying the Maxwell-Boltzmann statistics. In
this case of distinguishable classical atoms, the number of scattered
photons follow the fermionic system at high temperatures, as it should, but
starts to deviate at low temperatures (around one half of $E_F$) where it
continues to increase at a higher rate. Scattering of short laser pulses on
a system of trapped atoms thus provides a useful method for detecting the
temperature and hence degree of degeneracy of a fermionic system.

\section{Acknowledgments}

We thank Eric Bolda, Bereket Berhane, and Brian Kennedy for fruitful
discussions. This work is supported by the U.S. Office of Naval Research
grant No. 14-97-1-0633 and by the NSF grant No. PHY-9722410. The computation
of this work is also partially supported by NSF through a grant for the
ITAMP at Harvard University and Smithsonian Astrophysical Observatory.
M.~L. acknowledges the support of the SFB 407 of the Deutsche
Forschungsgemeinschaft.

\appendix

\section{Coherent spectrum}

Starting from Eq. (\ref{spcoh}) we explicitly write down the triple sum
expression for the coherent spectrum as
\begin{eqnarray}
C_{{\rm coh}}({\vec k},\mu)&=&S_{{\rm coh}}(\varpi) \Big|\sum_{n_x, n_y,
n_z} N_{n_x,n_y,n_z} {\cal I}_{n_x}{\cal I}_{n_y}{\cal I}_{n_z}\Big|^2,
\label{spcoh2}
\end{eqnarray}
where we have defined
\begin{equation}
{\cal I}_{n_j}=\langle n_j|e^{-i\delta k_j\cdot R_j} |n_j\rangle.
\end{equation}
Since $R_j$ is the position operator in the $j$-component ($j=x,y,z$), the
above matrix element is simply the diagonal elements of the displacement
operator, $\langle n_j|{\cal D}(-i\delta k_j a)|n_j\rangle$. We can then write
this matrix element in terms of the Laguerre polynomials using the following
identity \cite{Glauber},
\begin{equation}
\langle n|{\cal D}(\xi)|m\rangle = \sqrt{\frac{m!}{n!}} e^{-|\xi|^2
/2}\xi^{n-m}{\cal L}_m^{n-m}(|\xi|^2)  \label{identity}
\end{equation}
to obtain
\begin{equation}
\langle n_j|{\cal D}(\xi_j)|n_j\rangle = e^{-|\xi|^2 /2}{\cal L}%
_{n_j}(|\xi_j|^2),
\end{equation}
with $\xi_j=-i\delta k_j a$. We can now re-write Eq. (\ref{spcoh2}) as
\begin{eqnarray}
C_{{\rm coh}}({\vec k},\mu)&=&S_{{\rm coh}}(\varpi)e^{-|\xi|^2}\Big| %
\sum_{n_x, n_y, n_z} N_{n_x,n_y,n_z}  \nonumber \\
&\times& {\cal L}_{n_x}(|\xi_x|^2) {\cal L}_{n_y}(|\xi_y|^2) {\cal L}%
_{n_z}(|\xi_z|^2) \Big|^2 .
\end{eqnarray}
where $\xi=(\xi_x,\xi_y,\xi_z)$. We now rearrange the order of summation to
obtain
\begin{eqnarray}
C_{{\rm coh}}({\vec k},\mu)&=&S_{{\rm coh}}(\varpi)e^{-|\xi|^2}\Big| %
\sum_{n}^{\infty} \sum_{n_x+n_y+n_z=n} N_{n_x,n_y,n_z}  \nonumber \\
&\times& {\cal L}_{n_x}(|\xi_x|^2) {\cal L}_{n_y}(|\xi_y|^2) {\cal L}%
_{n_z}(|\xi_z|^2) \Big|^2 .
\end{eqnarray}
Since for the spherical symmetric trap under consideration, the mean
occupation number $N_{n_x,n_y,n_z}$ only depends on the sum of $n_x$, $n_y$
and $n_z$ ($n=n_x+n_y+n_z$), we can take it out of the inner sum and express
it as
\begin{equation}
P(n)=\frac{{\sf z}e^{-\beta\omega_t n}}{1+{\sf z}e^{-\beta\omega_t n}}.
\end{equation}
We then have
\begin{eqnarray}
C_{{\rm coh}}({\vec k},\mu)&=&S_{{\rm coh}}(\varpi)e^{-|\xi|^2}\Big| %
\sum_{n}^{\infty} P(n) \sum_{n_x+n_y+n_z=n}  \nonumber \\
&\times& {\cal L}_{n_x}(|\xi_x|^2) {\cal L}_{n_y}(|\xi_y|^2) {\cal L}%
_{n_z}(|\xi_z|^2) \Big|^2.
\end{eqnarray}
Using the summation theorem of Laguerre polynomials \cite{Sum_thm} to
replace the inner sum with a single generalized Laguerre polynomial we
finally obtain for the coherent spectrum,
\begin{equation}
C_{{\rm coh}}({\vec k},\mu)=S_{{\rm coh}}(\varpi) \Big| \sum_{n}^{\infty}
P(n) {\cal L}_n^2(|\xi|^2)\Big|^2 e^{-|\xi|^2}.
\end{equation}

%
%
%
%
%% figure 1
\newpage
%
%
%% figure 2
%\begin{figure}[b]
%\centerline{\epsfig{file=figures/fig2.eps,width=8.5cm}}
%\caption{The same as Fig.~1 but at a temperature of
%$k_BT/E_F=0.60$.} \label{fig02}
%\end{figure}
%
%
%% figure 3
%\newpage
%
%
%% figure 4
\newpage
%
%
%% figure 5
%
%
%% figure 6
\newpage
%
%
%% figure 7
%
%
%
%% figure 8
\newpage
%
%
%% figure 9
%
%
%% figure 10
\newpage
%\begin{figure}[t]
%\centerline{\epsfig{file=figures/fig10.ps,width=8.5cm}}
%\caption{The total number of coherent (circles) and incoherent (squares)
%scattered photons as a function of temperature for ten thousand trapped
%fermionic atoms. Coherent scattering calculated for a Maxwell-Boltzmann
%system is plotted as triangles. A $10$ ps laser pulse width was used in the
%calculations.}
%\label{fig10}
%\end{figure}
%
%
%
%
%% figure 11
\begin{figure}[b]
\centerline{\epsfig{file=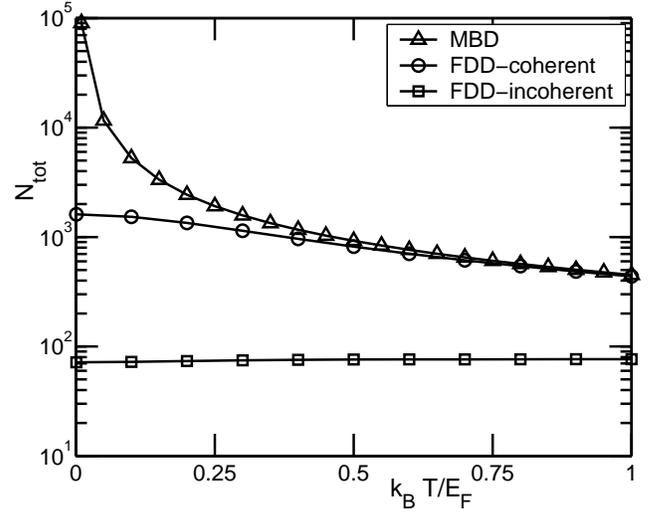,width=8.5cm}\\[12pt]}
\caption{The total number of coherent and incoherent scattered
photons as a function of temperature for one million trapped
fermionic atoms. Coherent scattering calculated for a
Maxwell-Boltzmann system is plotted as triangles. A $10$ ps laser
pulse width was used in the calculations.} \label{fig11}
\end{figure}

\end{document}